# Strong and weak separability conditions for two-qubits density matrices

Y. Ben-Aryeh

*Physics Department Technion-Israel Institute of Technology, Haifa, 32000, Israel*

E-mail: phr65yb@physics.technion.ac.il

ABSTRACT

Explicit separable density matrices, for mixed–two-qubits states, are derived by using Hilbert-Schmidt (HS) decompositions and Peres-Horodecki criterion. A "strongly separable" two-qubits mixed state is defined by multiplications of two density matrices, given with pure states, while "weakly separable" two-qubits mixed state is defined, by multiplications of two density matrices, which includes non-pure states. We find the sufficient and necessary condition for separability of two qubits density matrices and show that under this condition the two-qubits density matrices are strongly separable.

## 1. Introduction

For systems with many subsystems, and Hilbert spaces of large dimensions, the "separability problem" becomes quite complicated [1-2]. In the simple cases of two-qubits states, it is possible to give a measure of the degree of quantum correlations by using the partial-transpose (PT) of the density matrix [1-3]. According to Peres-Horodecki criterion [2-3], if the partial transpose of the two qubits density matrix leads to negative eigenvalues of the PT density-matrix $\rho_{AB}(PT)$, then the density matrix is entangled, otherwise it is separable.

One should take into account, that the density operator of a given mixture of quantum state has many ensemble decompositions. The separability problem for two-qubits states is defined as follows: A bipartite system is separable if the density matrix of this system can be transformed into the form:

$$\rho = \sum_j p_j \rho_A^{(j)} \otimes \rho_B^{(j)} \quad . \tag{1}$$

Here: $p_j \geq 0$, and $\sum_j p_j = 1$. The density matrix $\rho$ is defined on Hilbert space $H_A \otimes H_B$ where $A$ and $B$ are the two parts of a bipartite system. $\rho_A^{(j)}$, and $\rho_B^{(j)}$ are density matrices described, respectively, for the A and B systems. The interpretation for such definition is that for bipartite separable states the two systems, given by $\rho_A^{(j)}$, and $\rho_B^{(j)}$ are completely independent of each other. The summation over $j$ could include large numbers of density matrices multiplications, but it is preferred to limit this number to smaller ones, as far as it is possible.

The usual analysis of separability for two-qubits mixed states does not show, however, the explicit expressions for separable density matrices. I find the interesting distinction between "strong -separability", and "weak-separability". I define strong separability as given by Eq. (1) when

$$Tr\left(\rho^{(j)}_A\right)^2 = Tr\left(\rho^{(j)}_B\right)^2 = 1 \quad , \quad (j = 1, 2, \ldots) \tag{2}$$

Weak-separability is defined by Eq. (1) when some of the density-matrices $\rho_A^{(j)}$, and/or $\rho_B^{(j)}$ satisfy the relations

$$Tr\left(\rho^{(j)}_A\right)^2 < 1 \quad and/or \quad Tr\left(\rho^{(j)}_B\right)^2 < 1 \tag{3}$$

In more general terms conditions (2) and (3) are referred, respectively, as pure density matrices and mixed density matrices. The interesting point here is that while we assume $\rho$ to be a mixed state, the strong separability condition might still be valid. The explicit expressions for the density-matrices $\rho_A^{(j)}$, and $\rho_B^{(j)}$ might turn to be very complicated in general cases [4], but I restrict the discussion to two-qubits correlation density matrices which can be written in the Hilbert-Schmidt decomposition [5] as:

$$4\rho_{AB} = \left\{ (I)_A \otimes (I)_B + \sum_{i=1}^{3} t_i (\sigma_i)_A \otimes (\sigma_i)_B \right\} \tag{4}$$

Here $t_i$ ($i = 1, 2, 3$) are real parameters, $(\sigma_i)_A$ and $(\sigma_i)_B$ are Pauli matrices (i=1,2,3), $(I)_A$ and $(I)_B$ are $2 \times 2$ unit matrices, given, respectively, for the A and B subsystems. We described (4) in a frame in which the general matrix $t_{m,n}$ ($m, n = 1, 2, 3$) has a diagonal form. Under the symmetry condition: $t_{m,n} = t_{n,m}$, a straight forward transformation to the diagonal form (4) can easily be made. For the case for which $t_{m,n}$ is not symmetric $t_{m,n}$ matrix can be diagonalized by the use of singular value decomposition [10]. I find that the 2-qubits Bell states and the special Werner state analyzed in [6] are of this form. One should take into account: that $t_i$ ($i = 1, 2, 3$) are real parameters which can be both positive and negative. Also the multiplications in (4) are over Pauli matrices which are not density matrices. Our aim in the present article is to find relations between the two-qubit density matrix described by (4) and separable density matrices given by (1), and show that we get the strong separability condition although the total density matrix might be mixed.

## 2. Separability of two-qubits mixed states analyzed by Hilbert-Schmidt decomposition and Peres-Horodecki criterion

In the standard basis of states $|00\rangle, |01\rangle, |10\rangle, |11\rangle$, the density matrix (4) is given as:

$$4\rho_{AB} = \begin{pmatrix} 1+t_3 & 0 & 0 & t_1-t_2 \\ 0 & 1-t_3 & t_1+t_2 & 0 \\ 0 & t_1+t_2 & 1-t_3 & 0 \\ t_1-t_2 & 0 & 0 & 1+t_3 \end{pmatrix}. \tag{5}$$

The Partial-Transpose of the density matrix (5) is given by [7] as

$$4\rho_{AB}(PT) = \begin{pmatrix} 1+t_3 & 0 & 0 & t_1+t_2 \\ 0 & 1-t_3 & t_1-t_2 & 0 \\ 0 & t_1-t_2 & 1-t_3 & 0 \\ t_1+t_2 & 0 & 0 & 1+t_3 \end{pmatrix}. \tag{6}$$

The eigenvalues of $\rho_{AB}(PT)$ are given by:

$$4\lambda_1 = 1 + t_1 - t_2 - t_3 \quad , \quad 4\lambda_2 = 1 - t_1 + t_2 - t_3 \quad , \quad 4\lambda_3 = 1 - t_1 - t_2 + t_3 \quad , \quad 4\lambda_4 = 1 + t_1 + t_2 + t_3 \quad . \quad (7)$$

According to Peres-Horodecki criterion [1-3], if any one of the eigenvalues $\lambda_i$ $(i=1,2,3,4)$ is negative then the density matrix of (4) is entangled. I will give here explicit results for separability and entanglement as function of the absolute values of the constants: $t_i$. We distinguish between two cases:

**Case A: The sign of the triple product $t_1 t_2 t_3$ is -1 ( $sign(t_1 t_2 t_3) = -1$ ).**

We treat this case by 4 different conditions:

Condition a: If the three parameters $t_1, t_2, t_3$ are negative then the minimal value of $\lambda_i$ $(i=1,2,3,4)$ is given by

$$4\lambda_4 = 1 + t_1 + t_2 + t_3 = 1 - |t_1| - |t_2| - |t_3| \quad . \tag{8}$$

Condition b: If $t_1$ and $t_2$ are positive and $t_3$ is negative then the minimal value $\lambda_i$ $(i=1,2,3,4)$ is given by:

$$4\lambda_3 = 1 - t_1 - t_2 + t_3 = 1 - |t_1| - |t_2| - |t_3| \quad . \tag{9}$$

Condition c: If $t_1$ and $t_3$ are positive and $t_2$ is negative then the minimal value $\lambda_i$ $(i=1,2,3,4)$ is given by:

$$4\lambda_2 = 1 - t_1 + t_2 - t_3 = 1 - |t_1| - |t_2| - |t_3| \quad . \tag{10}$$

Condition d: If $t_2$ and $t_3$ are positive and $t_1$ is negative then the minimal value $\lambda_i$ $(i=1,2,3,4)$ is given by:

$$4\lambda_1 = 1 + t_1 - t_2 - t_3 = 1 - |t_1| - |t_2| - |t_3| \quad . \tag{11}$$

In order to get entanglement at least one of the eigenvalues should be negative.

We find that for cases with $sign(t_1 t_2 t_3) = -1$, the condition for entanglement is given by:

$$|t_1| + |t_2| + |t_3| > 1 \quad . \tag{12}$$

We have found according to Peres-Horodecki criterion for two-qubits system, that this condition is both sufficient and necessary. On the other hand if $|t_1| + |t_2| + |t_3| \leq 1$, then we get separable states. These results are the same for each of the four conditions a, or b or c or d.

**Case B: The sign of the triple product $t_1 t_2 t_3$ is +1 ( $sign(t_1 t_2 t_3) = +1$ ).**

We notice that any exchange of $t_i$ (i=1, 2 or 3) with $t_j$ ($j \neq i$) is equivalent to a corresponding exchange of two eigenvalues. For simplicity of notation, we assume:

$$|t_1| \geq |t_2| \geq |t_3| \quad , \tag{13}$$

We redefined the subscripts so that (13) is satisfied. We treat also this case for different conditions:

Condition a: If the three parameters $t_1, t_2, t_3$ are positive then the minimal value of $\lambda_i$ $(i = 1, 2, 3, 4)$ is given by

$$4\lambda_3 = 1 - t_1 - t_2 + t_3 = 1 - |t_1| - |t_2| + |t_3| \quad . \tag{14}$$

Condition b: If the two parameters $t_1$ and $t_2$ are negative and $t_3$ is positive then the minimal value of $\lambda_i$ $(i = 1, 2, 3, 4)$ is given by

$$4\lambda_4 = 1 + t_1 + t_2 + t_3 = 1 - |t_1| - |t_2| + |t_3| \quad . \tag{15}$$

Condition c: If the two parameters $t_1$ and $t_3$ are negative and $t_2$ is positive then the minimal value of $\lambda_i$ $(i = 1, 2, 3, 4)$ is given by

$$4\lambda_1 = 1 + t_1 - t_2 - t_3 = 1 - |t_1| - |t_2| + |t_3| \quad . \tag{16}$$

Condition d: If the two parameters $t_2$, and $t_3$, are negative and $t_1$ is positive then the minimal value of $\lambda_i$ $(i=1,2,3,4)$ is given by

$$4\lambda_2 = 1 - t_1 + t_2 - t_3 = 1 - |t_1| - |t_2| + |t_3| \qquad . \tag{17}$$

We find that if $sign(t_1 t_2 t_3) = +1$ under the condition

$$|t_1| + |t_2| - |t_3| > 1 \quad , \tag{18}$$

the density matrix becomes entangled. It has been shown [10], however, that a necessary and sufficient condition for the density matrix to be entangled, also in case B, is given by (12). For both cases the separable density matrices can become strongly entangled.

3. **Explicit expressions for the density matrices $\rho_A^{(j)}$, and $\rho_B^{(j)}$ of Eq. (1), for separable two qubits density matrices given by the density matrix (4), corresponding to cases A and B**

We will analyze the explicit expressions for separable density matrices which will be given here for the two cases derived above, by Peres-Horodecki criterion. We discuss the corresponding conditions for entanglement.

**Case A**: We study here the transformation of the density matrix (4) to a separable density matrix, given as a special case of (1), under the condition $|t_1| + |t_2| + |t_3| \leq 1$, for case A, defined by the relation $sign(t_1 t_2 t_3) = -1$. Such transformation will breakdown when $|t_1| + |t_2| + |t_3| > 1$, so that the results will be in agreement with the previous analysis made in Sec.2, by the use of Peres-Horodecki criterion. In order to find the entanglement properties of the density matrix (4), for case A: we define a matrix $S_{AB}$ by:

$$4S_{AB} = \sum_{i=1}^{3} t_i (\sigma_i)_A \otimes (\sigma_i)_B + (I)_A \otimes (I)_B \sum_{i=1}^{3} |t_i| . \tag{19}$$

The matrix $S_{AB}$ has the following properties:

a) $4\rho_{AB} - 4S_{AB} = \left[(I)_A \otimes (I)_B\right]\left(1 - \sum_{i=1}^{3}|t_i|\right)$ . (20)

Here $\rho_{AB}$, and $S_{AB}$ have been defined, respectively, in (4) and (19). The right side of (20) represents a separable density matrix (up to normalization) under the condition: $\sum_{i=1}^{3}|t_i| \leq 1$ where $\left(1 - \sum_{i=1}^{3}|t_i|\right)$ can be considered as a probability, but such representation breaks down when $\sum_{i=1}^{3}|t_i| > 1$, as we cannot have a negative probability.

b) $S_{AB}$ can be transformed into a form similar to (1) by using the following transformation:

$$4S_{AB} = \sum_{i=1}^{3} 2|t_i| \left[\left\{\frac{(I-\sigma_i)_A}{2} \otimes \frac{(I - sign(t_i)\sigma_i)_B}{2}\right\} + \left\{\frac{(I+\sigma_i)_A}{2} \otimes \frac{(I + sign(t_i)\sigma_i)_B}{2}\right\}\right]$$ (21)

$$= \sum_{i=1}^{3} 2|t_i|\left(\rho_i^{(-)} + \rho_i^{(+)}\right)$$

We defined here for positive $t_i$ (negative $t_i$) $sign(t_i) = 1$ $(sign(t_i) = -1)$. Each multiplication in the curled brackets on the right side of (21) represents a pure separable density matrix. We have defined, $\rho_i^{(-)}$ and $\rho_i^{(+)}$ (i=1, 2, 3) as the multiplication terms in the first and second curled bracket of (19). $\rho_i^{(+)}$, and $\rho_i^{(-)}$ are pure density matrices as they satisfy $Tr\left[\left(\rho_i^{(+)}\right)^2\right] = Tr\left[\left(\rho_i^{(-)}\right)^2\right] = 1$ $(i = 1, 2, 3)$ . $2|t_i|$, can be considered as a probability for each pure separable density matrix.

We can complement (21) with (20), obtaining the matrix $\rho_{AB}$ by

$$4\rho_{AB} = 4S_{AB} + \left[(I)_A \otimes (I)_B\right]\left(1 - \sum_{i=1}^{3}|t_i|\right)$$ . (22)

We have shown here how $\rho_{AB}$ of (1), in case A, is a separable density matrix with the use of pure density matrix multiplications, under the condition:

$$\sum_{i=1}^{3}|t_i|\leq 1 \qquad . \qquad (23)$$

The most interesting point is that we get explicit separable density matrics, by superposition of pure density matrices ("strong separability").

**Case B:** We study here the transformation of the density matrix (4) to a separable density matrix, given as a special case of (1). We studied the eigenvalues of (4) for case B defined by the condition: $sign(t_1 t_2 t_3) = +1$ , so that the results will be in agreement with the previous analysis made in Sec.2, by the use of Peres-Horodecki criterion. The condition $|t_1|+|t_2|-|t_3|>1$, is only a sufficient condition for entanglement. We find that sufficient and necessary condition for entanglement is given also for case B by $|t_1|+|t_2|+|t_3|>1$ [10]. I find that strong separability can be obtained also for case B . The separable density matrices can be given again by (21) where $|t_i|$ are considered as probabilities. The agreement between the separable density matrices, and (4) can be obtained by adapting the $sign(t_i)$ notations so that agreement will be obtained. We have shown that under the sufficient and necessary condition $|t_1|+|t_2|+|t_3|>1$ for entanglement for both cases A and B strongly separable states are obtained. Although I have treated various separability problems in previous articles [8-9], I have not analyzed there the explicit form of the separable density matrices.

### 4. Summary, discussion and conclusion

In the present work separability and entanglement properties of mixed two-qubits states have been analyzed by using Hilbert-Schmidt (HS) decompositions and Peres-Horodecki criterion. We have used a special form of two-qubits density matrix given by (4), depending on three diagonal constants $t_i$ $(i=1,2,3)$ . I have found that the eigenvalues of the two-qubits density matrix depend on the plus or minus sign of $t_1 t_2 t_3$. For the case of minus sign of this multiplication, referred in the article as case A, we obtained the condition $(|t_1|+|t_2|+|t_3|)>1$ for entanglement, while for the case with the plus sign for the above multiplication, referred in the article as case B, we get sufficient condition for entanglement given by $(|t_1|+|t_2|-|t_3|)\leq 1$ $(|t_1|\geq |t_2|\geq |t_3|)$. In order to get both sufficient and necessary conditions for entanglement one gets again for case B the same condition for entanglement $(|t_1|+|t_2|+|t_3|)>1$ as for case A [10]. These results follow from rigorous analysis of various cases by Pere-Horodecki criterion.

Explicit expressions for separable density matrices have been obtained by (21) and (22) for case A. For case B (21) and (22) can be used for describing separable density matrices by adapting the signs of $t_i$ given by the notations $sign(t_i)$. Although we analyzed special ensemble decompositions, these results seem to be quite general for these two cases. It is interesting to note that although we treat mixed two-qubits density matrices their decomposition for separable states included multiplications of two pure density matrices defined hare as strong separability conditions.